# Evaluating Grid Strength with Rising Penetration of Inverter Based Resources

**M. N. KHAMEES, G. ILUNGA, Z. ZHANG, R. E. TUCK**
**Dominion Energy Virginia**
**United States of America**

## SUMMARY

Over the years, Dominion Energy Virginia (DEV) has encountered recurring power quality disturbances, including voltage and power oscillations, particularly in proximity to inverter-based generation facilities. While these phenomena suggest a potential link to system strength, no definitive correlation has yet been established.

This paper aims to develop methodologies for assessing system strength across the DEV network and identifying regions potentially vulnerable to such disturbances. To achieve this, we propose calculating three key metrics to quantify grid strength—the Simple Short-Circuit Ratio (SSCR), the Weighted Short-Circuit Ratio (WSCR), and the Composite Short-Circuit Ratio (CSCR).

Historical event data will be analyzed using steady-state snapshots from DEV's proprietary Analysis on Demand (ANODE) platform Analysis on Demand (ANODE) platform, which provides 10-minute interval load flow data. These snapshots will be processed using the developed tool to compute system strength metrics. The analysis will explore correlations between system strength and observed disturbances. Where correlations are identified, we will further evaluate the potential impact of deploying Synchronous Condensers (SynCons) as a mitigation strategy.

mkhamees@vols.utk.edu

## INTRODUCTION

As inverter-based resources (IBRs) are increasingly integrated into power systems around the world, conventional synchronous generators powered by fossil fuels are being progressively retired. This ongoing technological transition introduces new and complex challenges for power system operation.

One of the key concepts used to study the interaction between the grid and power system equipment is grid strength. A strong grid is characterized by high voltage stiffness, meaning it can tolerate variations and disturbances with minimal impact. In contrast, weak grids present significant challenges for the integration of IBRs. These challenges include widespread undamped voltage and power oscillations, more severe voltage dips, higher transient overvoltages, reduced fault ride-through capability of generators, and potential malfunction or failure of protection systems.

Moreover, IBRs typically contribute little to grid strength, especially in areas where such support is most needed. Network sparsity plays a critical role in this context; IBRs connected to weak and sparse networks are particularly vulnerable to power quality issues [1]. Some IBR control schemes rely on terminal voltage measurements to synchronize with the grid. In weak systems, where voltage is highly sensitive to disturbances, this can lead to a loss of synchronism. Consequently, power systems with low system strength exhibit reduced resilience to voltage disturbances. In one short-circuit study, results for six scenarios indicated that the worst-case condition for composite short-circuit ratio (CSCR) and weighted short-circuit ratio (WSCR) was observed when an additional 125% increase of renewable energy was added, and all the main conventional generator units were out of service. [2].

To evaluate system strength, the Short-Circuit Ratio (SCR) is commonly used. The SCR is defined as the ratio between the short-circuit capacity at the bus where a device is connected and the device's rated power, measured in megawatts (MW). However, SCR does not account for the interactions between nearby renewable energy sources, which can lead to overly optimistic assessments of grid strength in cases where IBRs are electrically close. To address this limitation, enhanced metrics such as the Weighted Short-Circuit Ratio (WSCR) [3] and the Composite Short-Circuit Ratio (CSCR) [4] have been developed. These methods include the effects of interactions among adjacent renewable plants, providing a more accurate representation of system strength in high-IBR environments.

This paper seeks to precisely evaluate system strength in the Dominion Energy Virginia network that contains both conventional generators and renewable energy sources, with a focus on understanding how system strength relates to power quality concerns. The analysis centers on the effects of increasing renewable energy penetration by calculating the SCR, CSCR, and WSCR. Additionally, the study investigates how the integration of SynCons influences grid strength. The structure of the paper is as follows: Section II outlines the methodology; Section III describes the system setup and presents the system strength results, including the impact of growing renewable integration. Finally, Section IV concludes the study.

## METHODOLOGY

Voltage data recorded by the PingThings platform was analyzed to investigate historical power quality disturbances at substations where IBRs are connected. Focus was placed on the voltage measurements at the points of interconnections (POIs) where IBRs are connected, which helped trace the propagation of voltage oscillations and identify which IBRs dynamically oscillated together during events. To theoretically determine which IBRs are electrically close, we used electrical distance metrics derived from both impedance-based- and voltage sensitivity-based calculations.

The SCR at each POI was defined as the ratio of the Short-Circuit Capacity (SCC), expressed in MVA, to the rated power output (P) of the IBR at that location, measured in MW. The SCC is determined by applying a three-phase fault at the IBR's POI bus and measuring the fault current contribution from all non-IBR system components. To consider the interacting IBRs the CSCR was computed as follows:

$$CSCR = \frac{SCC}{\sum_{i}^{N} P_{MWi}} \quad (1)$$

where $SCC$ is the short-circuit capacity at bus $i$ without IBR contribution, $P_{MWi}$ is the rated output power of $IBR_i$, $N$ is the number of IBRs fully interacting with each other and $i$ is the IBR index. This method ties all IBRs of interest together by creating a common medium voltage bus.

When IBRs are not fully interactive but are still electrically close, the WSCR method can be applied to account for their influence:

$$WSCR = \frac{\sum_{i}^{N} SCC_i * P_{MWi}}{\sum_{i}^{N} {P_{MWi}}^2} \quad (2)$$

where $SCC_i$ is the short circuit capacity at bus $i$ without $IBR_i$ contribution, and $P_{MWi}$ is the rated output power of $IBR_i$. The CSCR and WSCR methods are typically applied when a more accurate evaluation of system strength is needed, especially in areas with multiple IBRs in close proximity. These methods offer precision over the conventional SCR approach.

After calculating grid strength across multiple events and locations, we will analyze the correlation between these events and grid strength levels. If a consistent relationship is observed, particularly when grid strength falls below a defined threshold, we will further investigate the potential benefits of adding a SynCon in the affected region.

SynCons are widely recognized for their effectiveness in strengthening power systems by boosting short-circuit capacity. Beyond enhancing system strength, they offer several operational advantages, including high overload tolerance, robust reactive power support during low-voltage conditions, and operation free from harmonic distortion. When examining the relationship between grid strength and disturbances, the deployment of SynCons is utilized as a mitigation strategy to address grid strength challenges. However, to ensure that their deployment meets dynamic performance requirements, it is essential to conduct a dynamic simulation study using detailed models.

## CASE STUDIES

Power flow scenarios were developed in PSSE using ANODE to perform short-circuit studies aimed at evaluating the effectiveness of power system strength assessment indices in IBR-integrated networks. The CSCR and WSCR metrics were computed as part of this analysis. IBRs were modeled with high source impedance to limit their short-circuit current contributions. Automation scripts were created to run three-phase fault simulations at key buses across the system.

**Case study 1:**

In this study, Dominion Energy's transmission network includes 34 IBRs. Figure 1 illustrates voltage oscillations observed at the POI of an IBR 3 where oscillations first reported. Among these, 12 IBRs were identified as being electrically close, indicating potential interactions and dynamic coupling between them. Voltage oscillations observed at the 12 IBRs.

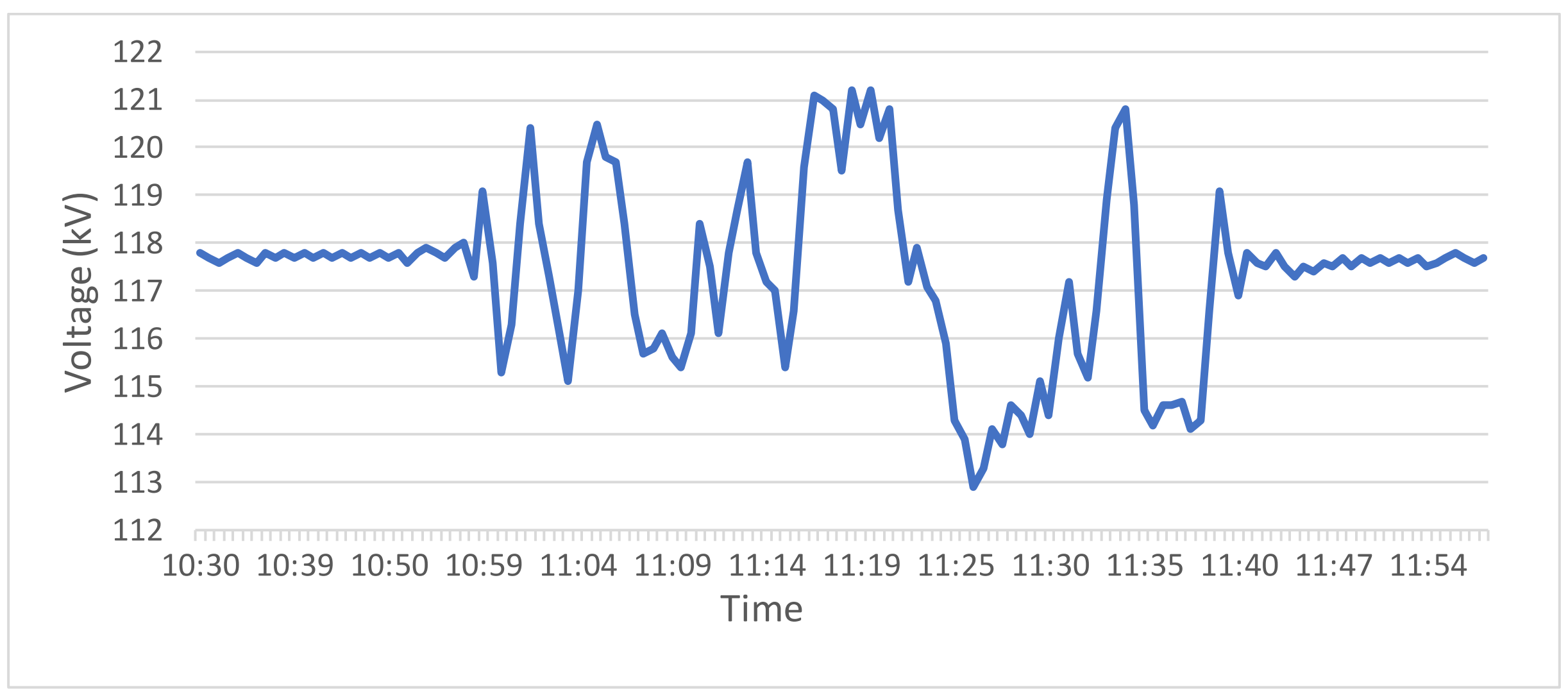

**Fig. 1. Voltage oscillation at POI of IBR 3.**

Table 1 presents the results of the short-circuit studies and the grid-strength metrics discussed in the previous section for the electrically close buses. As shown, the SCR values at all locations exceeded 9, indicating a strong system. However, a single SCR value is insufficient to capture the interactions among multiple IBRs. This highlights a limitation of using SCR as a standalone metric for system strength in IBR-dominated networks. To address this, WSCR and CSCR were employed to provide a more comprehensive assessment. While CSCR tends to overestimate system strength, computed at 5.75, assuming all IBRs are connected at a single bus, particularly in terms of voltage stability, WSCR offers a more conservative estimate. In this case, WSCR was calculated to be 1.47, below the commonly accepted minimum threshold of 1.5 for ensuring voltage stability.

TABLE 1
SYSTEM STRENGTH METRICS

| IBR | 1 | 2 | 3 | 4 | 5 | 6 | 7 | 8 | 9 | 10 | 11 | 12 |
|---|---|---|---|---|---|---|---|---|---|---|---|---|
| SCC (MVA) | 1023 | 1309 | 1518 | 1075 | 1409 | 1409 | 1650 | 1585 | 1609 | 685 | 1513 | 1608 |
| Capacity (MW) | 64.3 | 80 | 80 | 80 | 54 | 100 | 95 | 74 | 75 | 74 | 79 | 79 |
| SCR | 15.9 | 16.4 | 18.9 | 13.4 | 26.1 | 14.1 | 17.2 | 21.4 | 21.5 | 9.3 | 19.2 | 20.4 |

**Case study 2:**

In a separate incident, voltage fluctuations were detected at the POI of IBR 4, accompanied by oscillations in the same 12 IBRs mentioned in Case Study 1. The minimum SCR occurred at 11.82. The WSCR and CSCR were calculated to be 1.466 and 4.99, respectively. These results mirror the observations from Case Study 1.

**Case study 3:**

In a third scenario involving the network with 39 IBRs, seven were found to be electrically close to each other. The WSCR and CSCR values were 3.55 and 8.62, indicating a robust system overall. However, the SCR at the location where oscillations originated was 4.80, the lowest among all, suggesting a locally weak system.

To enhance the grid strength SynCons, which are synchronous machines without a prime mover, are commonly used to provide voltage support. In this study, a 50 MVar SynCon was installed at the weakest point identified in each of the three case studies to evaluate its impact. The results showed a noticeable improvement in system strength across all cases. In Case 1 and Case 2, the WSCR increased to 1.55 and 1.554, respectively. In Case 3, the lowest SCR rose significantly to 11.5. These outcomes demonstrate the effectiveness of SynCons in boosting SCC and mitigating weak grid conditions. However, to ensure comprehensive and efficient support across different IBR sites, it is recommended to conduct an optimal sizing and placement study for SynCons.

## CONCLUSION

The performance of IBRs is influenced by overall system strength. It is well known that integrating large-scale wind into a weak grid can lead to voltage oscillations and power quality issues. SCR is commonly used as an index for system strength, but it often ignores interactions between IBRs, which can result in an inaccurate or inflated assessment. This work applies to the concept of WSCR, which incorporates the full interaction effects among IBRs. WSCR offers a more realistic measure of system strength, especially when many wind units are connected to a weak grid. Nevertheless, further studies are needed to evaluate the accuracy of WSCR in capturing dynamic interactions among multiple IBRs. Based on the observations from case studies, it is advisable to perform an optimal sizing and placing of SynCon.